\documentclass[twocolumn,aps,prl]{revtex4-2}

\usepackage{amsmath, amssymb, physics, bm}
\usepackage{graphicx}

\begin{document}
\title{Universal behavior of the scattering matrix near thresholds in photonics}
\author{Charles C. Wojcik$^1$}
\author{Haiwen Wang$^2$}
\author{Meir Orenstein$^3$}
\author{Shanhui Fan$^1$}\email[Corresponding author: ]{shanhui@stanford.edu}
\
\affiliation{$^{1}$Department of Electrical Engineering, Ginzton Laboratory, Stanford University, Stanford, CA 94305, USA}
\affiliation{$^{2}$Department of Applied Physics, Ginzton Laboratory, Stanford University, Stanford, CA 94305, USA}
\affiliation{$^{3}$Department of Electrical Engineering, Technion-Israel Institute of Technology, Haifa 32000, Israel}
\date{\today}
 
\begin{abstract}
Scattering thresholds and their associated spectral square root branch points are ubiquitous in photonics. In this Letter, we show that the scattering matrix has a simple universal behavior near scattering thresholds. We use unitarity, reciprocity, and time-reversal symmetry to construct a two-parameter model for a two-port scattering matrix near a threshold. We demonstrate this universal behavior in three different optical systems, namely a photonic crystal slab, a planar dielectric interface, and a junction between metallic waveguides of different widths.
\end{abstract}

\maketitle


Scattering theory is the mathematical framework for describing physical systems with inputs and outputs. Originally developed to study the scattering of subatomic particles in quantum mechanics~\cite{taylor2006scattering, newton2013scattering} and to design microwave circuits~\cite{pozar2011microwave}, it has more recently become quite important in photonics~\cite{ebbesen1998extraordinary, astilean2000light, takakura2001optical, fan2002analysis, Fan:2003, suh2004temporal, shen2007strongly, ruan:2010, Chong:2010, verslegers2010temporal, Wan:2011, ge2012conservation, feng2014demonstration, kruk2017functional, sheinfux2017observation, zhao:2019} describing processes where the inputs and outputs are optical waves. 

The main object in scattering theory is the scattering matrix (S matrix), which relates the output amplitudes to the input amplitudes. The S matrix has a rich analytic structure which has been used to understand very general behavior of scattering processes. For example, poles of the S matrix have been used to develop a coupled-mode theory description of the Fano resonance in optical resonators~\cite{fan2002analysis, Fan:2003, suh2004temporal} and in the basic modal description of waveguides~\cite{chew1995waves}, while zeros of the S matrix have been used to design coherent perfect absorbers~\cite{Chong:2010, Wan:2011} and reflectionless scattering modes~\cite{sweeney2019theory}.


Besides poles and zeros, another universal analytic feature of the S matrix is the square root branch point~\cite{taylor2006scattering}. This branch point occurs when parameters such as frequency or angle are varied so that propagating channels (those with real phase accumulation) transition into evanescent channels (having imaginary phase accumulation). We refer to the transition point as a scattering threshold and the channels undergoing this transition (having zero phase accumulation) as threshold channels. For far-field engineering, the S matrix can be restricted to include propagating channels only, in which case the size of the S matrix changes at the threshold. However, for near-field engineering, the S matrix needs to be extended to include evanescent channels, in which case the size of the S matrix remains constant~\cite{eleftheriades1994some, carminati:2000}. In either case, the square root branch point appears as a distinctive spectral feature of the S matrix. 

In photonics, this square root singularity has been observed in the context of the Rayleigh-Wood's anomalies in the spectra of diffraction gratings and photonic crystals, associated with the appearance of new diffraction orders. It is well established that the origin of this square root behavior is the dispersion relation which determines the wavevector in terms of other parameters like frequency or angle~\cite{fano1941theory, hessel1965new}. There have also been detailed studies of the modified behavior of resonant poles in the presence of scattering thresholds~\cite{wigner1948behavior, lomakin2005enhanced, lomakin2006transmission, navarro2011negative, maurel2014wood, sheinfux2014subwavelength}. In the absence of resonant poles, the square root singularity has recently been considered as a possible means of enhanced sensing~\cite{sheinfux2017observation, chen2017exceptional}. Despite the theoretical and practical importance of understanding scattering thresholds, to the best of our knowledge there has not been a systematic treatment of the behavior of the S matrix at isolated scattering thresholds in photonics.  


In this Letter, we show that the square root branch point is associated with strong constraints on the entries of the S matrix, and thus that the S matrix has a simple universal behavior near scattering thresholds. Specifically, we argue on general grounds that at the scattering threshold, the threshold channels see a reflection coefficient of $-1$ and vanishing transmission coefficients (eqn.~\ref{eqn:m1}). Specializing to the two-port case and imposing the further constraints of unitarity, reciprocity, and time-reversal symmetry, we obtain a general two-parameter analytic model for the S matrix near a scattering threshold (eqn.~\ref{eqn:model}).

\begin{figure*}[t]
    \centering
    \includegraphics[width=0.99\textwidth]{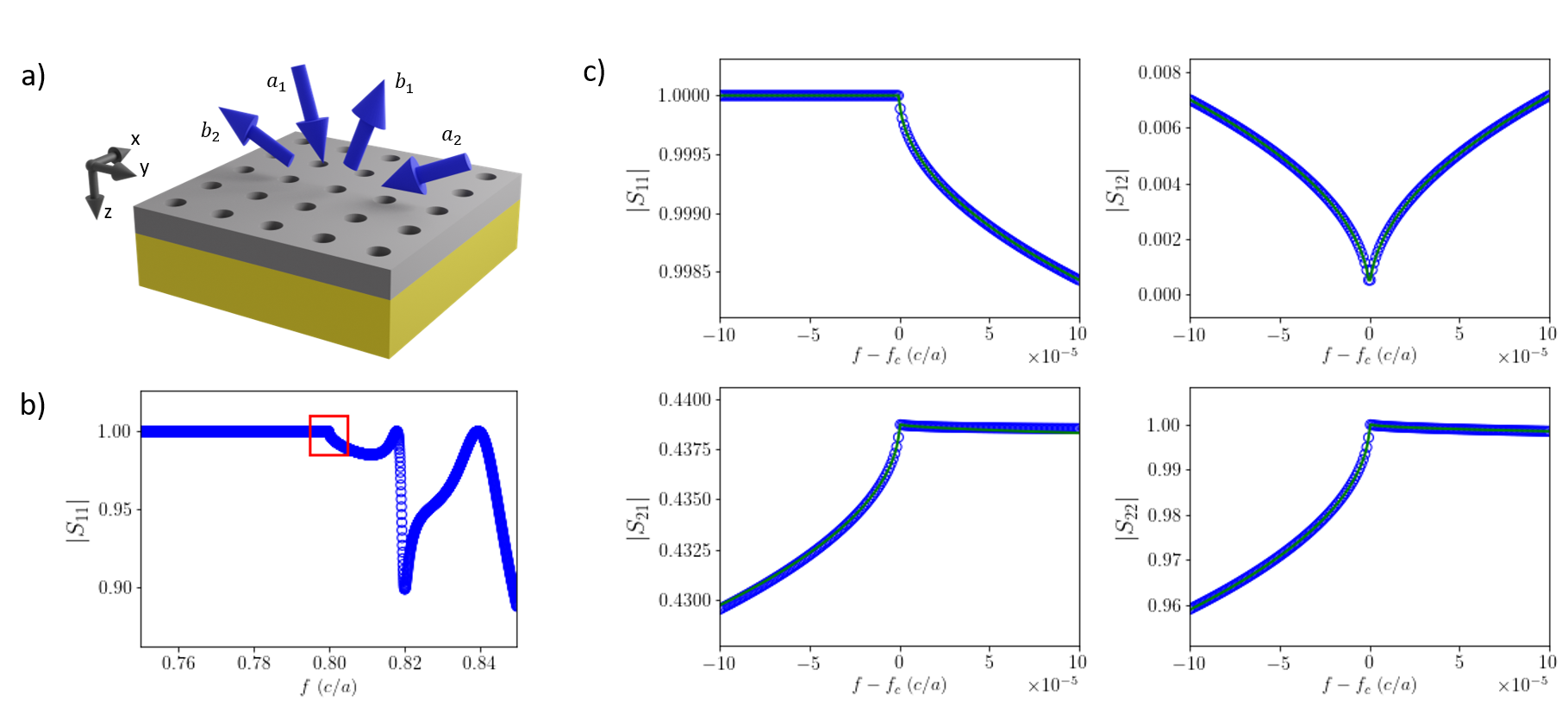}
    \caption{A scattering threshold in a photonic crystal slab. (a) The structure is a photonic crystal slab with dielectric constant $\epsilon = 3.5$ sitting on a perfect mirror. The air holes have radius $0.2a$, where $a$ is the lattice constant, and the slab thickness is $0.5a$. The arrows indicate the two scattering channels coupled by the structure, $i = 1,2$, each of which has an incoming $(a_i)$ and outgoing $(b_i)$ wave. The in-plane wavevector of channel $1$ is fixed to $k_x = 0, k_y = 0.2(2\pi/a)$, and the electric field is pointing in the $x$ direction (transverse electric, TE). (b) A portion of the reflection spectrum from channel $1$ is shown, with a red box indicating the kink at the scattering threshold. The critical frequency for the scattering threshold is $f_c = 0.8 c/a$, where $c$ is the speed of light. (c) The magnitude of the matrix elements of the two-channel S matrix near the threshold are shown. The circles are from simulations using RCWA~\cite{Jin2020} while the solid curve is from the two-parameter analytical model (eqn.~\ref{eqn:model}). The parameter values for the model (defined below) are $t = 0.438$ and $g = -1.282$. The primary features of the spectrum are the kink as well as the values $S_{12} = 0$ and $|S_{22}| = 1$ at threshold.}
    \label{fig:fig1}
\end{figure*}

We start with a numerical demonstration of the Rayleigh-Wood's anomalies in the spectrum of a photonic crystal slab. (In this and other examples, we consider transverse electric (TE) modes; equivalent results can be shown for transverse magnetic (TM) modes, but this is not shown for brevity.) This serves as a familiar example of a scattering threshold in photonics and illustrates some of the general features of the S matrix. We then turn to a theoretical analysis of the S matrix near a scattering threshold, deriving the general constraints and the two-parameter model. This analysis is motivated by the example of the photonic crystal slab, but we emphasize the generality of the arguments so that the results hold for arbitrary scattering thresholds. Finally, to illustrate the universality of these results, we look at two other systems with threshold behavior, namely a dielectric interface, where the threshold occurs at the critical angle for total internal reflection (TIR), and a junction between two metallic waveguides of different widths, where the threshold occurs at the cutoff frequency of the fundamental TE mode. In all cases we find good agreement with the two-parameter analytic model.

As a first example of a system with a scattering threshold, we consider a photonic crystal slab sitting on top of a perfect mirror, as shown in Fig.~\ref{fig:fig1}(a). We calculate the S matrix in this system using rigorous coupled wave analysis (RCWA)~\cite{Jin2020}. The photonic crystal slab couples two free-space scattering channels, $i = 1, 2$, each supporting an incoming ($a_i$) and outgoing $(b_i)$ wave. As we approach the critical frequency of the scattering threshold from above, the wavevectors associated with channel $2$ approach an angle which is parallel to the slab. Below the critical frequency, channel $2$ is evanescent in the $\beta$ direction (the $z$ component of the wavevector is imaginary). The $z$ component of the propagation constant associated with channel $2$ is
\begin{align*}
\beta &= 2\pi \sqrt{\bigg(\frac{f}{c}\bigg)^2 - \bigg(\frac{0.8}{a}\bigg)^2}.
\end{align*}
This is the origin of the threshold square-root branch point: the threshold critical frequency $f_c = 0.8 c/a$ for this diffraction order is defined by
\begin{equation*}
\beta(f = f_c) = 0
\end{equation*}
so that near the critical frequency
\begin{align*}
\beta \approx \frac{2\pi}{c} \sqrt{2 f_c(f - f_c)}.
\end{align*}
The S matrix is analytic as a function of $\beta$, so the square root branch point in this case comes from this relation between $\beta$ and $f$. In general, we will use the dispersion relations of the scattering channels to identify the critical values of the parameters associated with scattering thresholds.

In Fig.~\ref{fig:fig1}(b), we plot the magnitude reflection coefficient $|S_{11}|$ (which relates $b_1$ to $a_1$) as a function of frequency. We see a flat response below the critical frequency, since the second channel is evanescent and all of the power is reflected into the first channel. Above the critical frequency, there are two propagating channels, and so we see a typical photonic crystal spectrum with Fano resonances on a smooth background. Near the critical frequency, we see the square-root spectral kink (indicated by a red box) which is characteristic of a scattering threshold (in this case, the Rayleigh-Wood's anomaly).

In Fig.~\ref{fig:fig1}(c), we show a closer view of the magnitude of each entry of the two-port S matrix near the critical frequency. The notable features are the square-root kink at the threshold, as well as the fact that $S_{12} = 0$ and $|S_{22}| = 1$ at threshold. As we shall see, these features are universal at scattering thresholds. The numerical data shows a good fit with the two-parameter analytic model which we derive below (eqn.~\ref{eqn:model}).

Now that we have seen an example of a scattering threshold, we can provide a general analytic argument for the notable features. We consider a scattering system with $N$ channels, each of which can be propagating, evanescent, or threshold.
The electric and magnetic fields in the $n$th channel ($1 \leq n \leq N$) can be expressed in terms of forward and backward travelling waves as
\begin{align*}
(E_n)_x(z) &= a_n e^{-i \beta_n z} + b_n e^{i \beta_n z} \\
(H_n)_y(z) &= \frac{\beta_n}{\omega \mu_0} (a_n e^{-i \beta_n z} - b_n e^{i \beta_n z}).
\end{align*}
Here, $\beta_n$ is the propagation constant (the $z$ component of the wavevector), $\omega$ is the angular frequency, and $\mu_0$ is the vacuum permeability.

We now simplify the notation by evaluating at $z = 0$ and dropping the $x$ and $y$ subscripts on the electric and magnetic fields. We switch to a vector notation to treat all the channels at once, so $\mathbf{E}$ and $\mathbf{H}$ are $N$-component vectors, and we define the characteristic impedance matrix
\begin{align*}
\mathbf{Z_0} = \mathrm{diag}\bigg(\frac{\omega \mu_0}{\beta_n}\bigg).
\end{align*}
We can now write more simply
\begin{align*}
\mathbf{E} &= \mathbf{a} + \mathbf{b}  \\
\mathbf{H} &= \mathbf{Z_0}^{-1} (\mathbf{a} - \mathbf{b}).
\end{align*}
In addition to the characteristic impedance matrix, we also need the wave impedance matrix (Z matrix), defined by
\begin{align*}
\mathbf{E} &= \mathbf{Z} \mathbf{H}
\end{align*}
(recalling that $\mathbf{E}$ and $\mathbf{H}$ are vectors containing only transverse field components). Noting also the definition of the S matrix by the equation $\mathbf{b} = \mathbf{S} \mathbf{a}$, it follows that
\begin{align*} 
\mathbf{S} &= -(\mathbf{1} + \mathbf{Z} \mathbf{Z_0}^{-1})^{-1} (\mathbf{1} - \mathbf{Z} \mathbf{Z_0}^{-1})
\end{align*}
where $\mathbf{1}$ denotes the identity matrix. See Supplemental Material for the derivation. We later use this equation to derive a local two-parameter model near a threshold.

We now turn our attention to the behavior exactly at the scattering threshold. We partition the channels (and therefore the matrices $\mathbf{S}$ and $\mathbf{Z}$) into three blocks, corresponding to propagating channels, threshold channels, and evanescent channels respectively. The propagation constants of the threshold channels vanish at the critical frequency, so in our case (TE) the characteristic impedance of these channels is infinite. (This can be avoided in metamaterials where $\mu = 0$~\cite{reshef2019nonlinear}). The wave impedance (including cross-impedance between channels), on the other hand, generically remains finite at the threshold. From these properties, we can show that exactly at the threshold, the S matrix takes the block form
\begin{align}\label{eqn:m1}
\mathbf{S} &= \begin{pmatrix}
\mathbf{*} & \mathbf{0} & \mathbf{*} \\
\mathbf{*} & \mathbf{-1} & \mathbf{*} \\
\mathbf{*} & \mathbf{0} & \mathbf{*}
\end{pmatrix}
\end{align}
(where $-\mathbf{1}$ denotes the negative identity matrix, while asterisks denote entries which are not constrained by the threshold behavior). See Supplemental Material for the derivation. In other words, the transmission from any threshold channel to any other channel (including other threshold channels) is $0$, while the reflection coefficient from any threshold channel back to itself is $-1$.

To understand this behavior physically, we focus on the case where the input channel is a threshold channel. In our setup, at the threshold, the diverging characteristic impedance causes the magnetic field to go to zero. The finite wave impedance therefore requires that the electric field also vanishes; this requires the transmission coefficients $t_{mn} = S_{mn} (m \neq n)$ and reflection coefficients $r_n = S_{nn}$ to satisfy, when channel $n$ is exactly at threshold,
\begin{align*}
t_{mn} &= 0 \\
r_n &= -1.
\end{align*}

Now we develop a local model for the two-port S matrix. To the lowest order, we can assume that $\mathbf{Z}$ is constant in frequency, so that the analytic behavior of $\mathbf{S}$ comes entirely from the square root in $\mathbf{Z_0}$. Using unitarity, reciprocity, and time-reversal symmetry, we can show that $\mathbf{Z}$ is an imaginary symmetric matrix, which is therefore determined by three real parameters. See Supplemental Material for the derivation. The S matrix is therefore also determined by three real parameters. One of them is not physically relevant, since it corresponds to a choice of a reference plane for the propagating channel. In terms of the other two parameters, which we call $t$ and $g$, the S matrix has the form
\begin{align}\label{eqn:model}
\mathbf{S} &= \frac{1}{1 + (ig + (t/2)^2)\delta}  \nonumber \\
& \times \begin{pmatrix}
1 + (ig - (t/2)^2)\delta
& t \delta \\
t
& -1 + (ig + (t/2)^2) \delta
\end{pmatrix} 
\end{align}
where the dimensionless parameter $\delta$ is defined as the ratio of the two propagation constants,
\begin{equation*}
\delta = \beta_2 / \beta_1.
\end{equation*}
See Supplemental Material for the derivation. Recall that $\delta$ has a square-root frequency dependence near the threshold. The parameter $t$ is physical, and depends on the details of the scattering system. The parameter $g$ corresponds to a choice of reference plane for the threshold channel, but we keep it in the model because this choice of reference plane does affect the modulus of the S matrix elements. This is because below the threshold, a choice of reference plane corresponds to an imaginary gauge transformation on an evanescent wave~\cite{okuma2020topological}. The imaginary gauge transformations only affect $|S_{21}|$ and $|S_{22}|$ below the threshold.

\begin{figure}[b]
    \centering
    \includegraphics[width=0.49\textwidth]{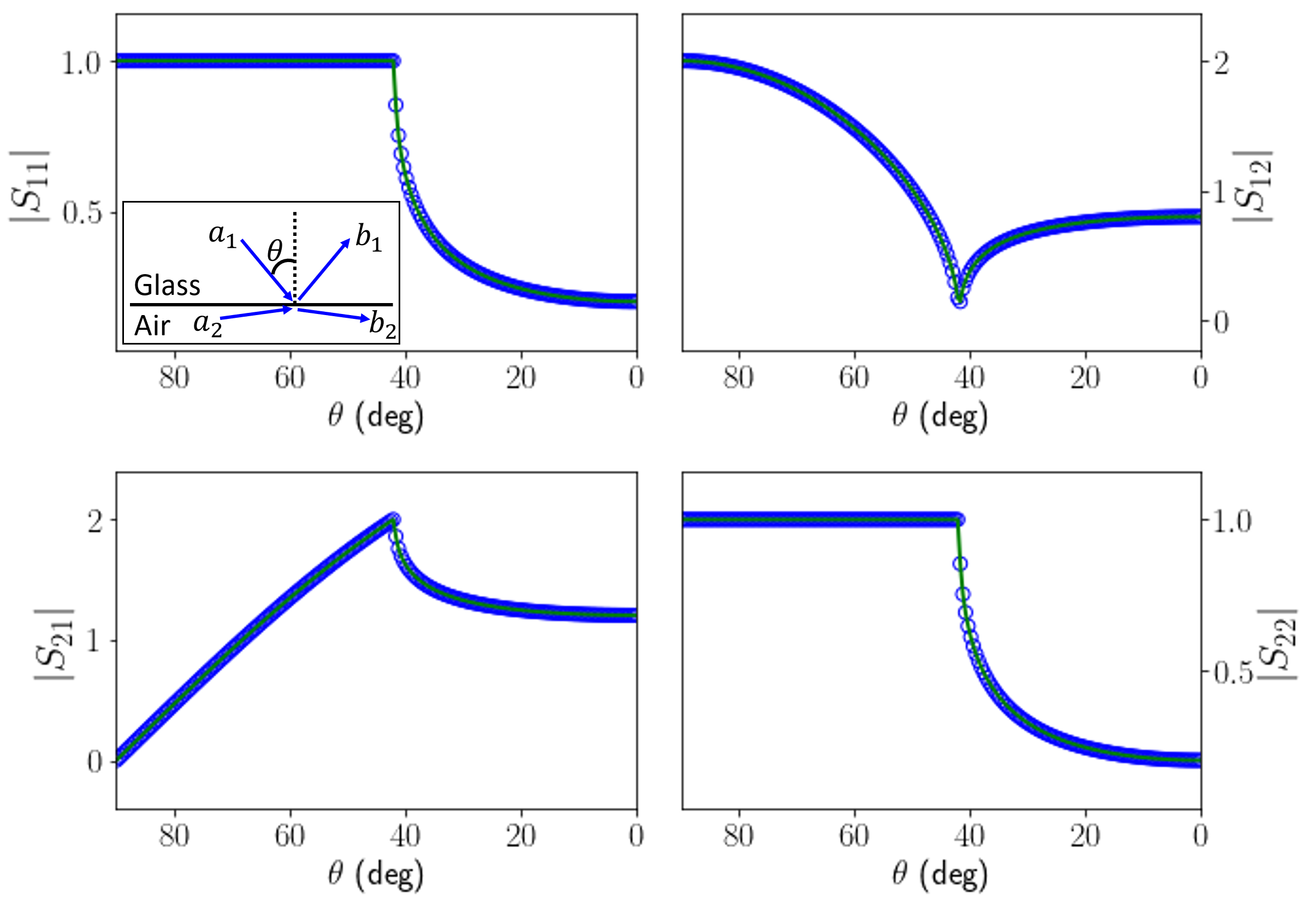}
    \caption{A scattering threshold at a dielectric interface. An inset shows the structure, a planar dielectric interface between glass (refractive index n = 1.5) and air (refractive index n = 1), as well as the angle of incidence and the two scattering channels. The four S matrix elements are plotted as a function of incident angle. The circles are from the Fresnel equations while the solid green curves are from the two-parameter analytical model (with $t = 2$ and $g = 0$). The scattering threshold occurs at the critical angle from TIR. The agreement in this case is exact for all angles.}
    \label{fig:fig2}
\end{figure}

Now that we have an analytic model for scattering thresholds in photonics, we illustrate the universality of this model by considering two other photonic systems with scattering thresholds. First, we consider the simplest system exhibiting threshold behavior, a planar interface between two different dielectric materials. This simple system is well-understood in terms of Snell's law of refraction, and the scattering threshold here occurs at the critical angle for TIR. For concreteness, we consider light incident from glass (refractive index $n_1 = 1.5$) onto an interface with air (refractive index $n_2 = 1$). The dispersion relation on the air side is
\begin{align*}
\beta_2 &= \frac{2 \pi f }{c} \sqrt{1 - (1.5 \sin \theta)^2}
\end{align*}
so we see that the threshold indeed occurs at the critical angle for TIR.
In Fig.~\ref{fig:fig2}, we show the configuration and the two scattering channels. We compare our model to the known Fresnel equations that describe transmission and reflection at the interface. In this simple system, the agreement with the model is exact. Accordingly, one can view the threshold from the Fresnel equations as characteristic of the local behavior at a general scattering threshold.

\begin{figure}[t]
    \centering
    \includegraphics[width=0.49\textwidth]{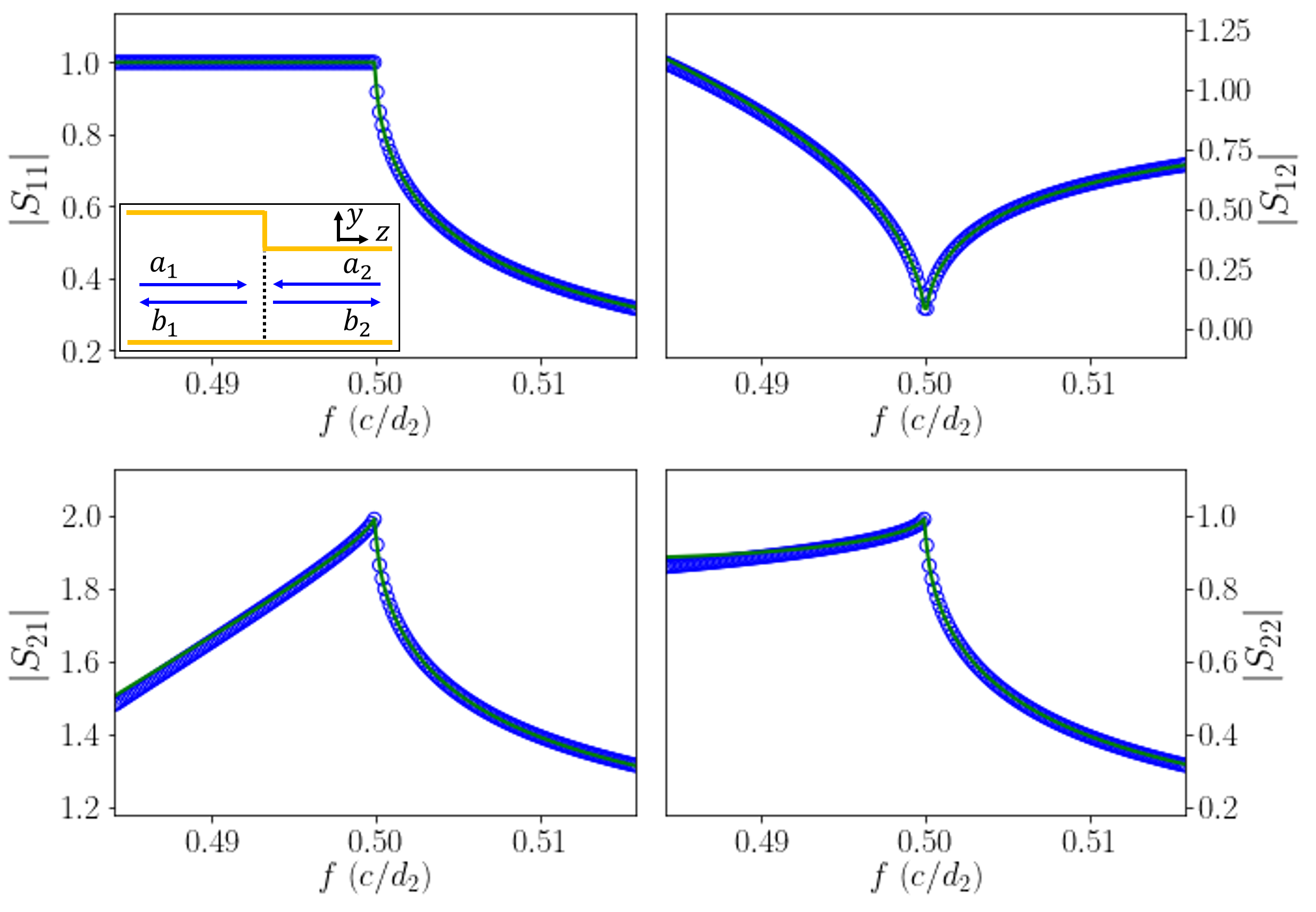}
    \caption{A scattering threshold at an junction between two metallic waveguides of different widths. An inset shows the structure; the first waveguide has width $d_1$ and the second waveguide has width $d_2$, where $d_1 > d_2$. Accordingly, the mode of the second waveguide has a scattering threshold at the critical frequency $f = 0.5 (c/d_2)$. The blue circles are numerical while the solid green curves are from the model (with $t = 2$ and $g = -0.127$).}
    \label{fig:fig3}
\end{figure}

For our last example, we consider a closed junction between two metallic waveguides of different widths, as shown in Fig.~\ref{fig:fig3}. The $n$th TE mode in a metallic waveguide of width $d$ has dispersion relation
\begin{equation*}
\beta_n = 2\pi\sqrt{\bigg(\frac{ f}{c}\bigg)^2 - \bigg(\frac{n}{2d} \bigg)^2}
\end{equation*}
so we see that the fundamental TE mode ($n = 1$) has a threshold at the cutoff frequency $f_c = 0.5 (c/d)$. In our system, the first and second waveguides have widths $d_1$ and $d_2$ respectively, and $d_1 > d_2$, so the fundamental TE mode in the first waveguide is still propagating at the cutoff frequency for the fundamental TE mode in the second waveguide. If $d_1$ is chosen small enough, the higher order TE modes in the first waveguide are not propagating. Because TE modes and TM modes do not couple in this system, we can again use our two-port analytic model to describe the S matrix of the fundamental TE modes in the two waveguides at a scattering threshold. See the Supplemental Material for the details of the analysis of this system~\cite{eleftheriades1994some, kocabacs2009modal}. Again, the two-parameter model predicts the critical behavior at the scattering threshold. 

Finally, we remark that the square root behavior and the constraints on the transmission and reflection coefficients at threshold are robust to the addition of loss to the scattering system, as long as the scattering channels themselves remain lossless (although the two-parameter model derived using unitarity breaks down). Of the example systems we considered here, this assumption holds in the photonic crystal slab (where the scattering channels are in air) but fails in the case of TIR and the metallic waveguide junction. We demonstrate this robustness for the photonic crystal slab with material loss in the Supplemental Material.

To summarize, we showed that the S matrix at a scattering threshold in photonics has a universal behavior and can be understood by a simple two-parameter model (eqn.~\ref{eqn:model}). The argument relies on the generic property that the characteristic impedance diverges while the wave impedance remains finite at thresholds. This universal behavior provides insight into widespread analytic features of the S matrix in photonics. The square root behavior of the S-matrix at threshold may also be important for sensing applications, and the analytic model would thus be useful for understanding the constraints for engineering these thresholds.

This work is supported by a Vannevar Bush Faculty Fellowship from the U. S. Department of Defense (Grant No. N00014-17-1-3030), and by the U. S. Office of Naval Research (Grant No. N00014-20-1-2450).




\begin{acknowledgments}

\end{acknowledgments}





\bibliography{bib}{}
\end{document}


\title{Supplemental Material: Universal behavior of the scattering matrix near thresholds in photonics}
\author{Charles C. Wojcik$^1$}
\author{Haiwen Wang$^2$}
\author{Meir Orenstein$^3$}
\author{Shanhui Fan$^1$}\email[Corresponding author: ]{shanhui@stanford.edu}
\
\affiliation{$^{1}$Department of Electrical Engineering, Ginzton Laboratory, Stanford University, Stanford, CA 94305, USA}
\affiliation{$^{2}$Department of Applied Physics, Ginzton Laboratory, Stanford University, Stanford, CA 94305, USA}
\affiliation{$^{3}$Department of Electrical Engineering, Technion-Israel Institute of Technology, Haifa 32000, Israel}
\date{\today}

\maketitle

\section{Relation between the $\mathbf{S}$ matrix and the $\mathbf{Z}$ matrix}
The scattering matrix $\mathbf{S}$ is defined in terms of forward and backward travelling wave amplitudes in each channel, which can be written as the vectors $\mathbf{a}$ and $\mathbf{b}$. Similarly, the impedance matrix $\mathbf{Z}$ is defined in terms of the electric and magnetic field amplitudes in each channel, which can be written as the vectors $\mathbf{E}$ and $\mathbf{H}$. The defining relations are
\begin{align*}
\mathbf{b} &= \mathbf{S} \mathbf{a} \\
\mathbf{E} &= \mathbf{Z} \mathbf{H}
\end{align*}
and we would like an equation relating the two matrices. We also know that
\begin{align*}
\mathbf{E} &= \mathbf{a} + \mathbf{b}  \\
\mathbf{H} &= \mathbf{Z_0}^{-1} (\mathbf{a} - \mathbf{b}).
\end{align*}
Substituting these equations for $\mathbf{E}$ and $\mathbf{H}$ into the defining equation for $\mathbf{Z}$, we obtain
\begin{align*}
\mathbf{a} + \mathbf{b} = \mathbf{Z} \mathbf{Z_0}^{-1} (\mathbf{a} - \mathbf{b}).
\end{align*}
We can solve this equation by
\begin{align*}
 \mathbf{b} &= -(\mathbf{1} + \mathbf{Z} \mathbf{Z_0}^{-1})^{-1} (\mathbf{1} - \mathbf{Z} \mathbf{Z_0}^{-1})\mathbf{a}
\end{align*}
so by the definition of the scattering matrix, we have
\begin{align}\label{eqn:cayley}
\mathbf{S} &= -(\mathbf{1} + \mathbf{Z} \mathbf{Z_0}^{-1})^{-1} (\mathbf{1} - \mathbf{Z} \mathbf{Z_0}^{-1}).
\end{align}

\section{Constraints on the $\mathbf{Z}$ matrix due to unitarity, reciprocity, and time-reversal invariance}

In this section, we will show that the impedance matrix $\mathbf{Z}$ as defined by
\begin{align*}
\mathbf{E} = \mathbf{Z} \mathbf{H}
\end{align*}
satisfies
\begin{align*}
\mathbf{Z}^\dag &= - \mathbf{Z} \\
\mathbf{Z} &= \mathbf{Z}^T
\end{align*}
and therefore
\begin{align*}
\mathbf{Z}^* &= - \mathbf{Z}.
\end{align*}
We will only explicitly check the unitarity and reciprocity constraints, since in general time-reversal invariance follows from these two.

First, consider the time-averaged Poynting vector, which in our notation is $S = \frac{1}{2}\Re{E_n H_n^*}$ in the $n$th channel. Since the energy is conserved when all channels are considered, we have
\begin{align*}
0 &= \sum_n \Re{E_n H_n^*} \\
&= \Re \langle \mathbf{E}, \mathbf{H} \rangle \\
&= \Re \langle \mathbf{Z}\mathbf{H}, \mathbf{H} \rangle \\
&= \frac{1}{2}(\langle \mathbf{Z}\mathbf{H}, \mathbf{H} \rangle + \langle \mathbf{H}, \mathbf{Z}\mathbf{H} \rangle) \\
&= \frac{1}{2}\langle (\mathbf{Z} + \mathbf{Z}^\dag)\mathbf{H}, \mathbf{H} \rangle.
\end{align*}
Since this holds for all $\mathbf{H}$, we must have $\mathbf{Z}^\dag = - \mathbf{Z}$.

Next, we look at the reciprocity constraint. We consider two independent sets of fields $(\mathbf{E}, \mathbf{H}), (\mathbf{\tilde{E}}, \mathbf{\tilde{H}})$, and we find that
\begin{align*}
0 &= \sum_n E_n\tilde{H}_n - \tilde{E}_n H_n \\
&= \mathbf{Z}\mathbf{H} \cdot \mathbf{\tilde{H}} - \mathbf{Z}\mathbf{\tilde{H}} \cdot \mathbf{H} \\
&= (\mathbf{Z} - \mathbf{Z}^T) \mathbf{H} \cdot \mathbf{\tilde{H}}
\end{align*}
so we conclude that $\mathbf{Z}^T = \mathbf{Z}$.

Finally, we note that using these constraints on $\mathbf{Z}$ as well as our equation relating $\mathbf{S}$ and $\mathbf{Z}$ provides a simple alternative derivation of the constraints on the S matrix in the presence of evanescent modes~\cite{carminati:2000}.

\section{S matrix exactly at a scattering threshold}
In this section, we calculate the S matrix exactly at a scattering threshold. We start with equation~\ref{eqn:cayley} to analyze the threshold behavior. At the threshold, we have seen that the characteristic impedance of the threshold channels are infinite, while the wave impedance matrix is finite. Therefore exactly at the threshold (as a block matrix)
\begin{align*}
\mathbf{1} + \mathbf{Z}\mathbf{Z_0}^{-1} &= \begin{pmatrix}
\mathbf{*} & \mathbf{0} & \mathbf{*} \\
\mathbf{*} & \mathbf{1} & \mathbf{*} \\
\mathbf{*} & \mathbf{0} & \mathbf{*}
\end{pmatrix}
\end{align*}
(using asterisks to denote entries which are not determined by the threshold behavior). It follows from Cramer's rule for matrix inversion~\cite{lang2012introduction} that the inverse of such a matrix has the same form (as a block matrix). The other factor, $(\mathbf{1} - \mathbf{Z} \mathbf{Z_0}^{-1})$, has the same form as well. It therefore follows that exactly at the threshold, the S matrix takes the block form
\begin{align*}
\mathbf{S} &= \begin{pmatrix}
\mathbf{*} & \mathbf{0} & \mathbf{*} \\
\mathbf{*} & -\mathbf{1} & \mathbf{*} \\
\mathbf{*} & \mathbf{0} & \mathbf{*}
\end{pmatrix}
\end{align*}

\section{Local model for two-channel S matrix}
In this section, we derive the local model for the two-channel S matrix near a threshold, again starting with equation~\ref{eqn:cayley}.
Write the impedance matrix at threshold as
\begin{align*}
\mathbf{Z} &= \begin{pmatrix}
a & b \\
c & d
\end{pmatrix}.
\end{align*}
For now, we don't assume any constraints on $a, b, c$, and $d$, but later, we will specialize using unitarity, reciprocity, and time-reversal symmetry to the case where $b = c$ and all of the parameters are pure imaginary. Write
\begin{align*}
\mathbf{Z_0}^{-1} &= \begin{pmatrix}
\frac{\beta_1}{\omega \mu_0} & 0 \\
0 & \frac{\beta_2}{\omega \mu_0}
\end{pmatrix}
\end{align*}
so
\begin{align*}
\mathbf{Z}\mathbf{Z_0}^{-1} &= \begin{pmatrix}
a\frac{\beta_1}{\omega \mu_0} & b\frac{\beta_2}{\omega \mu_0} \\
c\frac{\beta_1}{\omega \mu_0} & d\frac{\beta_2}{\omega \mu_0}
\end{pmatrix}.
\end{align*}
Rescaling $a, b, c$, and $d$ by $\frac{\beta_1}{\omega \mu_0}$ and writing $\delta = \beta_2 / \beta_1$, we have
\begin{align*}
\mathbf{Z}\mathbf{Z_0}^{-1} &= \begin{pmatrix}
a & b \delta \\
c & d \delta
\end{pmatrix}.
\end{align*}

Now we calculate (to lowest order in $\delta$)
\begin{align*}
\mathbf{S} &= -\begin{pmatrix}
1 + a & b\delta \\
c & 1 + d\delta
\end{pmatrix}^{-1}\begin{pmatrix}
1 - a & -b\delta \\
-c & 1-d\delta
\end{pmatrix} \\
&= -\frac{1}{(1 + a)(1 + d\delta) - bc\delta} \\
& \times \begin{pmatrix}
1 + d\delta & -b\delta \\
-c & 1 + a
\end{pmatrix}\begin{pmatrix}
1 - a & -b\delta \\
-c & 1 -d\delta
\end{pmatrix} \\
&= -\frac{1}{(1 + a)(1 + d\delta) - bc\delta} \\
& \times \begin{pmatrix}
(1 + d\delta)(1 - a) + bc\delta & -2b\delta \\
-2c & bc\delta + (1 + a)(1 - d\delta)
\end{pmatrix} \\
&= \frac{1}{1 + (d - \frac{bc}{1 + a})\delta} \\
& \times \begin{pmatrix}
-\frac{1-a}{1+a}[1 + (d + \frac{bc}{1-a})\delta] & \frac{2b}{1+a}\delta \\
\frac{2c}{1+a} & -1 + (d - \frac{bc}{1+a})\delta
\end{pmatrix}.
\end{align*}

At this point, we assume unitarity so that all of the parameters are pure imaginary. Write
\begin{align*}
\frac{2b}{1 + a} &= t_{12} e^{i \theta} \\
\frac{2c}{1 + a} &= t_{21} e^{i \theta}
\end{align*}
and note that
\begin{align*}
-\frac{1-a}{1+a} &= \Big( \frac{-2b}{1-a} \Big)^{-1} \frac{2b}{1 + a} \\
&= (t_{12} e^{-i \theta})^{-1} t_{12} e^{i \theta} \\
&= e^{2i\theta}.
\end{align*}
Furthermore,
\begin{align*}
\Re{\frac{bc}{1+a}} = \Re{\frac{bc}{1-a}} &= \frac{bc}{(1+a)(1-a)} \\
&= -t_{12} t_{21}/4 \\
\Im{\frac{bc}{1+a}} = -\Im{\frac{bc}{1-a}}
\end{align*}
so defining
\begin{align*}
g &= \Im{d - \frac{bc}{1+a}}
\end{align*}
we obtain
\begin{align*}
\mathbf{S} &= \frac{1}{1 + (ig + (t_{12} t_{21}/4))\delta} \\
&\times \begin{pmatrix}
e^{2i\theta}[1 + (ig - (t_{12} t_{21}/4))\delta]
& t_{12} e^{i\theta}\delta \\
t_{21} e^{i\theta}
& -1 + (ig + (t_{12} t_{21}/4)) \delta
\end{pmatrix}.
\end{align*}

\begin{figure}[t]
    \centering
    \includegraphics[width=0.49\textwidth]{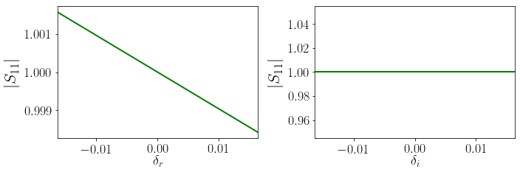}
    \caption{The magnitude $|S_{11}|$ is shown as a function of $\delta$ for $\delta = \delta_r$ purely real or $\delta = i \delta_i$ purely imaginary. The matrix element $S_{11}$ is an analytic function of $\delta$ at the critical point $\delta = 0$.}
    \label{fig:figS1}
\end{figure}

Clearly the $\theta$ parameter represents a choice of reference plane for the propagating channel, so we can omit it from our model if we only care about the amplitude of the scattering coefficients. Furthermore, if reciprocity holds, then we have $t_{12} = t_{21} = t$, so the S matrix takes the form
\begin{align*}
\mathbf{S} &= \frac{1}{1 + (ig + (t/2)^2)\delta} \\
&\times \begin{pmatrix}
1 + (ig - (t/2)^2)\delta
& t\delta \\
t
& -1 + (ig + (t/2)^2) \delta
\end{pmatrix}.
\end{align*}

To illustrate the analyticity of the S matrix as a function of $\delta$, in Fig.~\ref{fig:figS1} we show $|S_{11}|$ as a function of $\delta$ for $\delta$ either purely real or purely imaginary, i.e. $\delta = \delta_r$ or  $\delta = i \delta_i$. As in our first example of the photonic crystal slab from the main text, we take the parameters to be $t = 0.438$ and $g = -1.282$. We note that the square root dependence of $\delta$ on frequency is responsible for the square root branch point of the S matrix, and when the frequency passes through the critical frequency, $\delta$ changes from purely imaginary to purely real.

\section{Mode matching method for an interface between metallic waveguides}
Consider an interface between two metallic waveguides of widths $d_1$ and $d_2$ with $d_1 > d_2$. The fundamental TE modes in waveguide $i$ have a cutoff frequency $\omega_{c,i} = c \pi / d_i$ which is larger for waveguide $2$ than for waveguide $1$. For $\omega_{c,1} < \omega < \omega_{c,2}$, the mode in waveguide $1$ is propagating and the mode in waveguide $2$ is evanescent, so we can expect to see two-channel threshold behavior at the cutoff frequency $\omega_{c,2}$. The tangential components of the TE electric and magnetic fields have the modal expansion
\begin{align*}
E_x(y, z < 0) &= 
\sum_{n=1}^\infty \sqrt{\frac{2}{d_1}} \sin(\frac{n \pi y}{d_1}) \\
&\times (a_{n, 1} e^{-i \beta_{n, 1} z} + b_{n, 1} e^{i \beta_{n, 1} z}) \\
E_x(y, z > 0) &= 
\sum_{n=1}^\infty \sqrt{\frac{2}{d_2}} \sin(\frac{n \pi y}{d_2}) \\
&\times (a_{n, 2} e^{-i \beta_{n, 2} z} + b_{n, 2} e^{i \beta_{n, 2} z}) 
\\
H_y(y, z < 0) &= \sum_{n=1}^\infty \frac{\beta_{n, 1}}{\omega \mu_0} \sqrt{\frac{2}{d_1}} \sin(\frac{n \pi y}{d_1}) \\
&\times (a_{n, 1} e^{-i \beta_{n, 1} z} - b_{n, 1} e^{i \beta_{n, 1} z}) \\
H_y(y, z > 0) &= \sum_{n=1}^\infty \frac{\beta_{n, 2}}{\omega \mu_0} \sqrt{\frac{2}{d_2}} \sin(\frac{n \pi y}{d_2}) \\
&\times (a_{n, 2} e^{-i \beta_{n, 2} z} - b_{n, 2} e^{i \beta_{n, 2} z}).
\end{align*}
In these expressions, we have understood $\beta_{n, 2}$ to be negative so that we can use a uniform coordinate system for the counter-propagating waves in the two channels.

The relevant boundary conditions at the interface between the two waveguides are
\begin{align*}
E_x(y, z < 0) &= \begin{cases}
E_x(y, z > 0) & \textrm{ for } 0 < y < d_2\\
0 & \textrm{ for } d_2 < y < d_1
\end{cases} \\
H_y(y, z < 0) &=
H_y(y, z > 0) \textrm{ for } 0 < y < d_2.
\end{align*}
We also have the orthogonality relations
\begin{align*}
\int_{0}^{d_i} \sin(\frac{m \pi x}{d_i})\sin(\frac{n \pi x}{d_i}) \dd{x} &= \frac{d_i}{2} \delta_{mn}.
\end{align*}
Integrating the electric field boundary condition against the modes in the larger waveguide and the magnetic field boundary condition against the modes in the smaller waveguide~\cite{eleftheriades1994some, kocabacs2009modal} gives
\begin{align*}
a_{m, 1} + b_{m, 1} &= \sum_{n=1}^\infty Q_{mn} (a_{n, 2} + b_{n, 2}) \\
\beta_{m, 2}(a_{m, 2} - b_{m, 2}) &= \sum_{n=1}^\infty Q_{nm} \beta_{n, 1}(a_{n, 1} - b_{n, 1})
\end{align*}
where $\mathbf{Q}$ is the matrix whose entries are given by
\begin{align*}
Q_{mn} &= \frac{2}{\sqrt{d_1 d_2}} \int_{0}^{d_2} \sin(\frac{m \pi x}{d_1}) \sin(\frac{n \pi x}{d_2})\dd{x}.
\end{align*}
We note that $\mathbf{Q}^T \mathbf{Q} = 1$ on the Hilbert space formed by the modes of the smaller waveguide, while $\mathbf{Q} \mathbf{Q}^T$ is a projection onto the span of the modes of the smaller waveguide (the range of $\mathbf{Q}$) inside the Hilbert space of the larger waveguide. Define the diagonal matrices
\begin{align*}
\bm{\beta_i} = \mathrm{diag}(\beta_{n,i})
\end{align*}
and write $\mathbf{a_i}, \mathbf{b_i}$ to denote the vectors whose entries are $a_{n,i}, b_{n,i}$ respectively. Then we have the matrix equations
\begin{align*}
\mathbf{a_1} + \mathbf{b_1} &= \mathbf{Q}(\mathbf{a_2} + \mathbf{b_2}) \\
\bm{\beta_2}(\mathbf{a_2} - \mathbf{b_2}) &= \mathbf{Q}^T \bm{\beta_1}(\mathbf{a_1} - \mathbf{b_1}).
\end{align*}

These equations can be solved numerically after truncating to a finite number of modes in each waveguide. Alternatively, we can truncate to a single mode in each waveguide to obtain the approximate analytic solution
\begin{equation*}
S = \frac{1}{1 + Q^{-2} \delta}\begin{pmatrix}
1 - Q^{-2} \delta & 2 Q^{-1} \delta \\
2 Q^{-1} & -1 + Q^{-2} \delta
\end{pmatrix}
\end{equation*}
with $Q = Q_{11}$. We see that this lowest order approximation agrees exactly with the two-parameter analytic model if we take $t = 2Q^{-1}, g = 0$.


\section{Effect of loss}

\begin{figure}[b]
    \centering
    \includegraphics[width=0.49\textwidth]{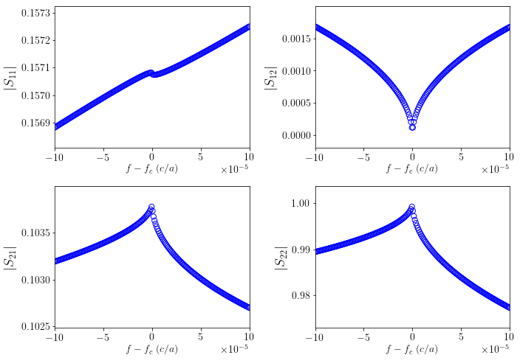}
    \caption{The S matrix elements are shown for the photonic crystal slab structure discussed in the main text, but this time with dielectric constant $\epsilon = 3.5+1j$. The square root behavior as well as the value of $S_{12}$ and $S_{22}$ at the threshold are unchanged, as expected from the general argument we give for these features. However, the two-parameter analytic model we derived no longer applies, as that derivation assumes unitarity.}
    \label{fig:figS2}
\end{figure}

To see which properties of the S matrix near threshold are robust to the addition of loss, in Fig.~\ref{fig:figS2} we show the photonic crystal slab structure from the main text but this time with the slab dielectric constant given by $\epsilon = 3.5+1j$. The square root branch point and the values of the reflection and transmission coefficients seen by the threshold channel that we predict on general grounds are robust. The two-parameter analytic model which we derived using unitarity no longer applies to this lossy system.

\bibliography{bib}{}